\documentclass[useAMS,usenatbib]{mn2e}
\usepackage{psfig, epsf, epsfig}
\begin{document}
\title[Bayesian approach to cyclic activity of CF Oct]{Bayesian approach to cyclic activity of CF Oct}
\author[A. P. Borisova, V. V. Hambaryan and J. L. Innis]{A. P. Borisova$^{1}$\thanks{E-mail: aborisova@astro.bas.bg (APB); vvh@astro.uni-jena.de (VVH); brightwater@iraf.net (JLI)} and V. V.
Hambaryan$^{2}$\footnotemark[1] and J. L. Innis$^{3}$ \footnotemark[1]\\
$^{1}$Institute of Astronomy, Bulgarian Academy of Sciences, 72 Tsarigradsko Shosse blvd., BG-Sofia, 1784, Bulgaria \\
$^{2}$Astrophysical Institute and University Observatory, 2-3 Schillerg\"{a}sschen, Jena, D-07745, Germany \\
$^{3}$Brightwater Observatory, 280 Brightwater Rd., Howden, TAS, 7054, Australia}
\date{Received 2010 August 04; last edition 2011 November 16}
\pagerange{\pageref{firstpage}--\pageref{lastpage}} \pubyear{2011}
\maketitle
\label{firstpage}

\begin{abstract}
Bayesian statistical methods of Gregory-Loredo and the Bretthorst generalization of the Lomb-Scargle periodogram have been applied for studying activity cycles of the early K-type subgiant star CF Oct. We have used a $\sim45$~year long dataset derived from archival photographic observations, published photoelectric photometry, Hipparcos data series and All Sky Automated Survey archive. We have confirmed the already known rotational period for the star of $20.16$~d and have shown evidences that it has exhibited changes from $19.90$~d to $20.45$~d. This is an indication for stellar surface differential rotation.The Bayesian magnitude and time--residual analysis reveals clearly at least one long-term cycle. The cycle lenght's posterior distributions appear to be multimodal with a pronounced peak at a period of $7.1$~y with $FWHM$ of $54$~d for time-residuals and at a period of $9.8$~y with $FWHM$ of $184$~d for magitude data. These results are consistent with the previously postulated cycle of $9\pm3$ years. 
\end{abstract}


\section{Introduction}
Long-term stellar activity study is a challenging research area so far as it needs the longest available data sets, and appropriate statistical methods to treat the problematic in its complexity.
\citet{philhart} first reported the long-term photometric variability of spotted binary stars, based on the archival photographic plate photometry combined with photoelectric photometric data. Their work produced more than 50 years of activity history for BY~Dra and CC~Eri. A similar result for V833~Tau was published by \citet{hart}, and is later confirmed by \citet{bon}. Star--spot activity is now known to occur on a range of solar--type and early K~giants stars. Some relevant works in the field are those of \citet{mess}, \citet{bon}, \citet*{olah1}, \citet{olahstrass}.

In the present analysis we used combined photographic, photoelectric photometry and CCD observations to study cyclic activity of the spotted subgiant star CF~Oct. The collected data, presented on Fig. 1, are quasi--randomly distributed with data gaps over time interval of about 45 years. We applied the Gregory-Loredo Bayesian method for time series with independent Gaussian noise, \citet{GL2}, in order to obtain Bayesian estimate for the known $\sim20$~d rotational modulation and to search for long-term periodic variability in time-scales up to about 15 years. The method and the IDL procedure, developed in the AIU Jena, that we used have been previously tested on the synthetic data-sets, modeled with known period and amplitude, observational errors and simulated data gaps. They have shown reliable results and accurate period estimation.

\section{CF Oct - active spotted giant star}
CF Oct (HD 196818, HIP 102803, ${\rm \alpha}=20^{h}49^{m}37^{s}.263$, ${\rm \delta}=-80^{\circ}08'01''.01$, J2000, $K0 IIIp$ given by \citet{houk}) is a bright, southern active giant star. Its variability was first detected by \citet{stroh} on the photographic plates from Bamberg Observatory Southern Sky Survey (BOSSS). In the GCVS \citep{sam} the star is described as a RS~CVn type variable with maximal brightness $V=8.27$~mag and relatively large photometric variations $\sim 0.3$~mag. Early archival observations from the BOSSS were recently digitized and analyzed by \citet{ib1}.

Photoelectric photometry for the star was performed and published by \citet{Innis1}, \citet{lloyd}, \citet{poll}, \citet*{innis2}. Rapid changes in the light curve in $2006$ was announced by \citet*{innis3}. The photometry studies reveal rotational modulation of $\sim{20}$~d, due to spotted activity. Spectroscopic studies by \citet{hear} and \citet{innis2} of CF Oct show strong Ca\,{\sc ii} emission and filled-in H$_{\alpha}$ line. \citet{innis2} reported for no significant radial velocity variations and hence concluded that the star is probably single. Based on the radial velocity data and the period of the rotational cycle they consider it is a subgiant rather than a giant as per Houk classification. CF~Oct is also reported to be a strong, flaring, microwave radio source by \citet{slee}, and appears at the ROSAT Bright survey catalogue \citep{fish} with 1.12 counts per second in the energy range of 0.1--2.4~keV.

\section{Observations}
For the present study we have used data from: the Bamberg Observatory Southern Sky Survey (BOSSS), published photoelectric photometry observations, the Hipparcos satellite time-series, and the All Sky Automated Survey (ASAS) data archive.

BOSSS \citep{tcvb} was taken in the period 1963 to 1976 and contains more than 22 000 plates that cover the whole southern sky with limiting magnitude in range $11$ to $14$~mag. This survey is unique because of the fact that in this time interval the first Palomar and Harvard Observatory Southern Sky Surveys were completed, but other southern observatories were not actively surveying the sky. CF~Oct was observed on 352 plates in the period 1964--1976. \citet{ib1} performed aperture photometry and phase dispersion minimization (PDM) light curve analysis. In \citet{ib1} we discussed the methods for photographic plate digitization, the plate photometry and the transformation from $B$ to $V$ magnitude of the star. Plate data (BAM) are collected in more then six seasons relatively well covered by observations, separated by significant intervals with a lack of observations.

Photoelectric photometry data (PHOT) came from following papers: \citet{Innis1}, \citet{lloyd}, \citet{poll}, \citet{innis2}. The dataset consists of 137 data points, that might be separated in 5 seasons of intense observations.
In this PHOT dataset the mean brightness of the star is at its historical minimum. It is not clear whether this is an observational selection effect. \citet{lloyd} applied systematic magnitude corrections in order to have agreement for the photometry from different seasons, although they noted this was purely an ad hoc approach. Changes of mean light level have been observed for other RS~CVn stars (\citet*{fro} for HK~Lac, \citet{poretti} for AY~Cet).

The Hipparcos and Tycho Catalogues \citet{esa}, available via the Centre de Données astronomiques de Strasbourg (CDS), provide high quality scientific data for the period November 1989--March 1993. For our research we have used the Hipparcos time series data, as they have relatively smaller errors and are close to the photometry from Tycho catalogues. The Hipparcos magnitudes of CF Oct were transformed to $V$ by the use of equations given by \citet{bess}.

The most recent photometry for our work is extracted from the ASAS, \citet{asas}, archive at the web-page http://www.astrouw.edu.pl/asas/. The ASAS project aims at photometric monitoring of the whole available sky, from two observing stations -- Las Campanas Observatory, Chile since 1997 and Haleakala, Maui since 2006. Both stations are equipped with two wide-field 200--mm/f/2.8 instruments, observing simultaneously in $V$ and $I$ band. The project's ultimate goal is detection and investigation of any kind of photometric variability. As CF Oct is a relatively bright star, so far the early ASAS photometry data are unusable due to saturated images. For this reason, after consulting with ASAS supporting team, for the analysis we included only the data free of saturation influence.

The dataset we have collected contains data for HJD of the observation, $V$ mag of the star, and corresponding errors. As far as it consists of four datasets taken from different sources it suffers from significant intervals with lack of data, and also by non-uniform data distribution. The resulting $V$ magnitude light curve with overploted errors is presented on Fig.~1, where the crosses present BAM data, asterisk - PHOT data, diamonds - HIP data and triangles - ASAS data. The data-set statistically is presented in Table~1, with following information: Dataset ; $N_{\rm p}$ - the Number of data points in the set; $T_{\rm s}$ - the time span of the set in days; HJD is in the beginning of the set; $V_{\rm min}$, $V_{\rm max}$ and $<V>$ - minimal, maximal and mean values of $V$ magnitudes respectively.

\begin{figure}
\psfig{file=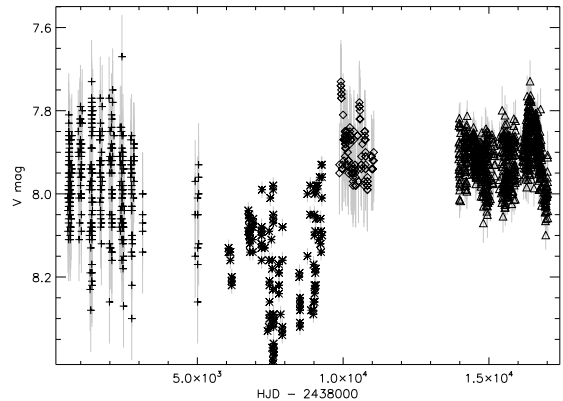,width=8.0cm}
\caption{
$V$ magnitude light curve of CF~Oct for the period 1964--2009 with overploted errors in gray colour.
}
\end{figure}

\begin{table}
 \centering
 \begin{minipage}{140mm}
 \caption{Statistical description of the data}
 \begin{tabular}{@{}llrrrrlrlr@{}}
 \hline
 Dataset&$N_{\rm p}$ & $T_{\rm s}$& $HJD$ & $V_{\rm min}$ & $V_{\rm max}$& $<V>$ \\
 \hline
BAM&352&4484&2438560&7.67&8.30&7.98 \\
PHOT&137&3212&2444071&7.93&8.41&8.16 \\
HIP&130&1176&2447873&7.74&7.98&7.90 \\
ASAS&705&3058 &2451961&7.73& 8.07&7.91\\
\hline
\end{tabular}
\end{minipage}
\end{table}

\section{Bayesian Statistical methods in use}
We used Gregory-Loredo method(GL), \citet{GL2}, for time-series analysis in the presence of Gaussian errors. It employs Bayes's theorem for detection of a periodic signal with unknown shape and period and for estimation of the characteristics of the detected signal. First step of the Bayesian statistical methods is to determine the hypothesis space. In general, for observed stars we have three hypotheses - constant light, variable light, and a periodic signal. In the GL method periodic models are represented by a signal folded into a stepwise function, similar to a histogram, with \textbf{$m$} phase bins per period plus a noise contribution. With such a model we are able to approximate a light curve of any shape.

The stepwise model is then described by the following parameters: period \textbf{$P$}, phase of the minimal star brightness \textbf{$\phi$}, number of bins \textbf{$m$}, light curve shape \textbf{$r_i$} in the each bin and noise scale parameter \textbf{$b$}. The GL method uses Jeffreys prior distributions for \textbf{$b$} and \textbf{$P$}, to ensure fully compatibility of the posterior distribution functions of the period and the frequency (\textbf{$1/P$}) , and uniform prior distribution for the rest of the parameters. The prior range for the period and the light curve shape parameter are user selected according the available data and the prior information.

The constant model may be treated just as a special case of a periodic model with only one bin \textbf{$m = 1$} and it has only two parameters: the value of the constant signal and the noise scale parameter. In the stepwise representation the nonperiodic modulation model may be treated as another special case of periodic model, when the period is equal to the data time span, with wider range of the number of bins and it has \textbf{$m$} light-curve shape parameters, bin phase and the noise scale parameter.

The stepwise model is flexible enough to approximate light curve of practically any shape, and although it is well suited to describe ''spiky'' signals, \citet{GL2} points out that it can successfully detect smooth signals and accurately estimates their shape and parameters. The posterior mean values, the most probable values and the credible intervals of the parameters are then estimated by marginalization of the global likelihoods over the priori specified range of each parameter. In our particular case the posterior mean is a robust estimator for the rotational period the of CF~Oct. Bayesian posterior probability contains a term that penalizes complex models, hence we calculate the posterior probability by marginalizing over a range of models, corresponding to a prior range of \textbf{$m$} from 2 to 12. The noise scale parameter \textbf{$b$} in GL method is the relation of the variances of the observer's errors and model residuals, and it is an indication whether the model successfully describes the observational data. If the mode of the \textbf{$b\approx{1}$}, then the proposed model is accounting for everything that is not noise in the data. If the mode of the \textbf{$b<1$}, then the model does not account significant features in the data, or the initial observers error estimates are low. Values of the \textbf{$b>1$} is an indication for over-fitting as the model residuals are smaller than the observational errors. We have tested the method by using modeled datasets, with randomly distributed data-points and with data gaps. For independent verification of the detected signals we have also applied Lomb-Scargle method, together with the Bayesian generalization of the Lomb-Scargle periodogram by \citet{bret}.

In order to reduce the likelihood of introducing unknown systematic errors into the periodicity study of CF~Oct, due to the different observational methods and data reduction procedures employed, instead of analyzing $V$ magnitude variations, we have analyzed the variations of $<V>-V$, where $<V>$ is the mean value of $V$ magnitude for each data set, i.e. $<\textbf{V}>$=($<V_{\rm bam}>$,$<V_{\rm phot}>$,$<V_{\rm hip}>$,$<V_{\rm asas}>$). On a short time-scale the star is already known as a variable, so while we analysed the rotational modulation we concentrate on testing the hypothesis for detecting periodic variations, and on the period estimation. With the collected data we are able to study variability of CF~Oct in ranges from several days (this limit is set up by the average sampling frequency of our observations) to 15 years (1/3 of the covered observational time span). This is a rather large time span, and was examined in several parts by using a suitable number of frequencies in each section. The long-term variability study of the star was performed by estimating the odds ratios (the ratio of the probabilities of the model with respect to the constant model) of the nonperiodic and periodic modulation models, as well as by searching for periodic modulation.

\section{Detecting the rotational modulation}

As CF Oct has a well established rotational period near~$\sim20$~d, known from PDM and least-squares periodical analysis of photographic and photoelectric observations, we have applied the Gregory-Loredo method for its detection. This is relevant as already studied BAM and PHOT data sets show slightly different periods \citep{ib1} that might be due to the period estimation errors, phase shifting or more complex periodical modulation. All the data sets are rich enough and are suitable to be studied separately as well as all together. For the separate datasets and for all the combined data we have calculated the global posterior probability density functions for a class of models described by the following parameters : \textbf{$b$} noise scale in the priori range from $0.05$ to $2.0$, number of bins (\textbf{$m$}) in range from 2 to 12; the period ($P$) in a prior range from $19$ to $21$~d, covered with relevant number of frequencies. We have followed \citet{GL2} in the case of $\textbf{$b$}\neq{1}$ and estimated the projected probability density function of the noise scale parameter. In case of the PHOT and ASAS data we have used a model with additional error term, in order to avoid numerical difficulties because of the very low observational errors.

\begin{table*}
 \centering
\begin{minipage}{90mm}

\caption{Derived parameters: number of bins, period and 68 per cent credible interval }
\begin{tabular}{@{}lllrrcc@{}}
 \hline
Dataset & \textbf{$m_{\rm mode}$} & \textbf{$b_{\rm mode}$} & \textbf{$P_{\rm mode}$} & \textbf{$P_{\rm mean}$} &\textbf{$cre. int., (from-to)$} &\textbf{$P_{\rm bret}$} \\
&Value & (days) & (days) & (days) &(days) &(days)\\
\hline
BAM&3&0.95&20.04&20.04&20.035--20.045&20.04 \\
PHOT&2&0.80&20.17&20.16&20.145--20.175&20.16 \\
HIP&2&1.00&20.46&20.45&20.425--20.475&20.51 \\
ASAS&3&0.90&19.94&19.94&19.930--19.950&19.93 \\
\hline
ALL&3&0.28&20.16&20.14&20.120--20.950&20.11 \\
\hline
\end{tabular}

\end{minipage}
\end{table*}

The parameters of interest are estimated by marginalization of the global likelihood function (equation 7, \citet{GL2}) over the nuisance parameters in the priori range. Marginalization over the noise scale parameter is the most conservative way for estimation of the model parameters that treats anything in the data, not described by the model as a noise. By marginalization of the joint posterior likelihood function over the frequencies we have computed the probability of the number of bins, and thus the most probable \textbf{$m$} (\textbf{$m_{\rm mode}$} in Table 2). The number of bins parameter relates to the complexity of the light curve to the shape of the light-curve and probably is connected with the structure of the stellar spots (or the spot groups). The values of the \textbf{$m_{\rm mode}$} parameter, we obtained, tend to be small 2 or 3 with a rather narrow distribution, so that model with one more light curve bin is already penalized by the Occam's razor factor for involving unreasonable number of parameters.
\begin{figure*}
\begin{center}
\vbox{
\hbox{
\psfig{file=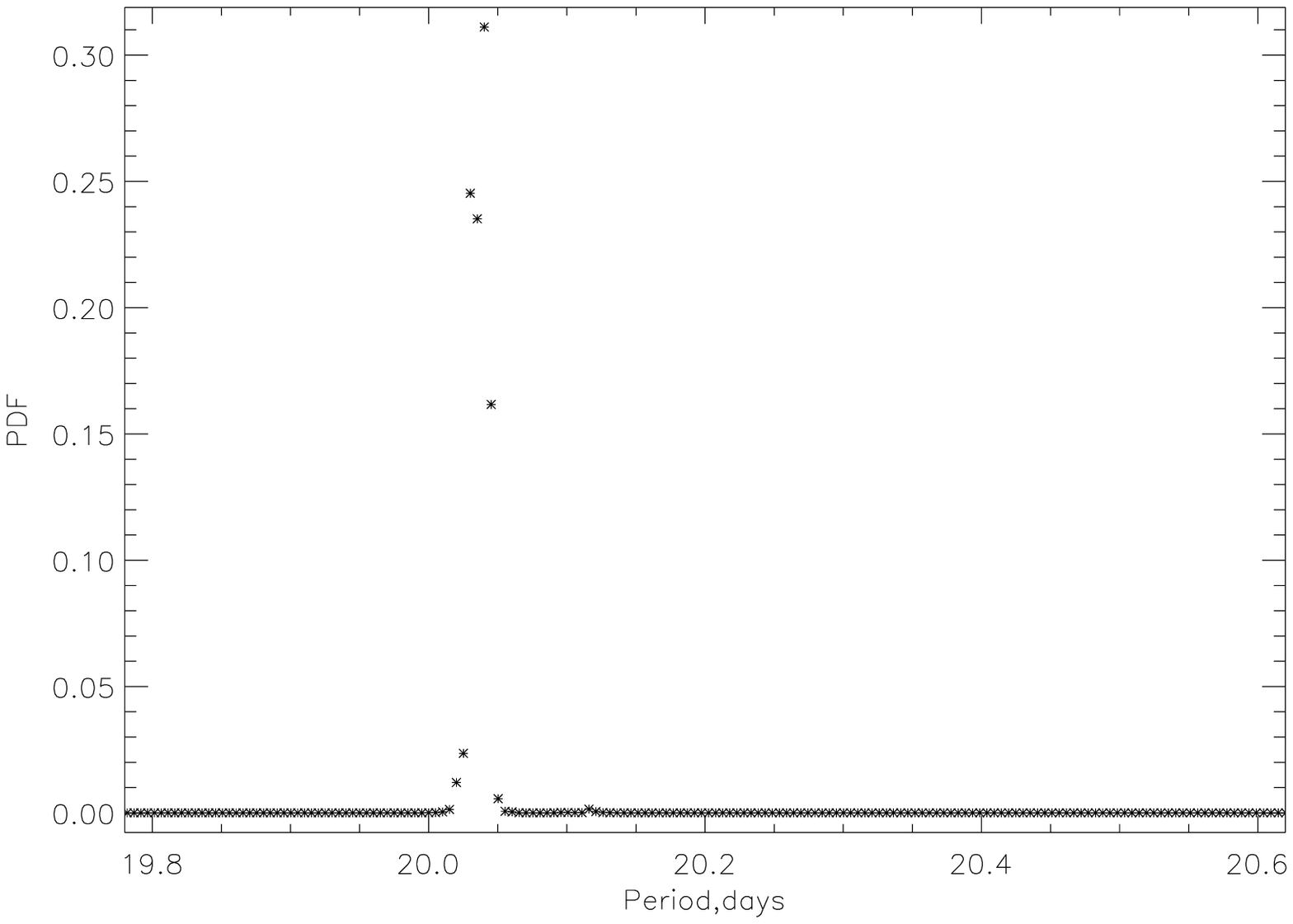,width=7.5cm, height=4.5cm}
\hspace{0.25cm}
\psfig{file=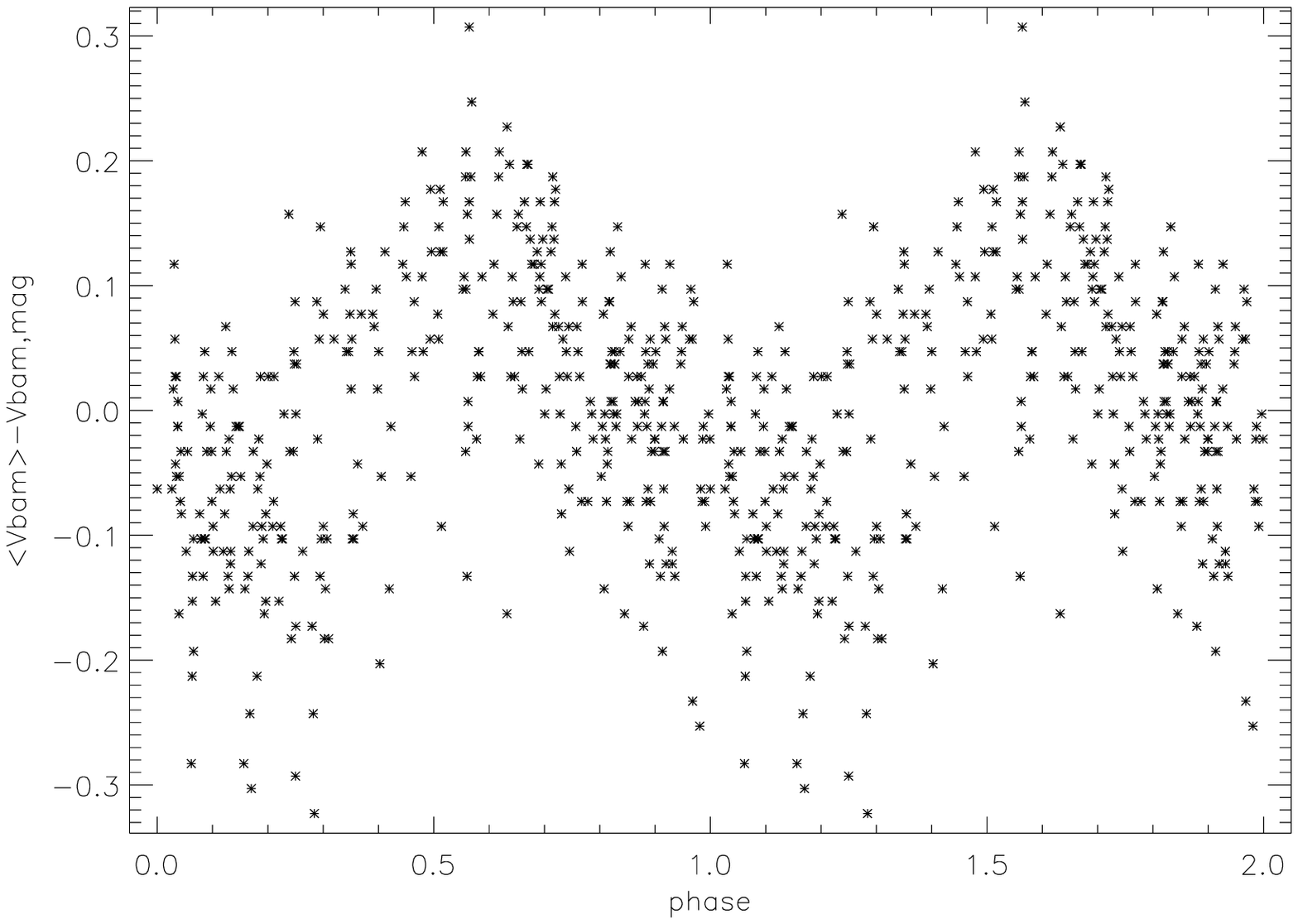,width=7.5cm, height=4.5cm}
}
\hbox{
\psfig{file=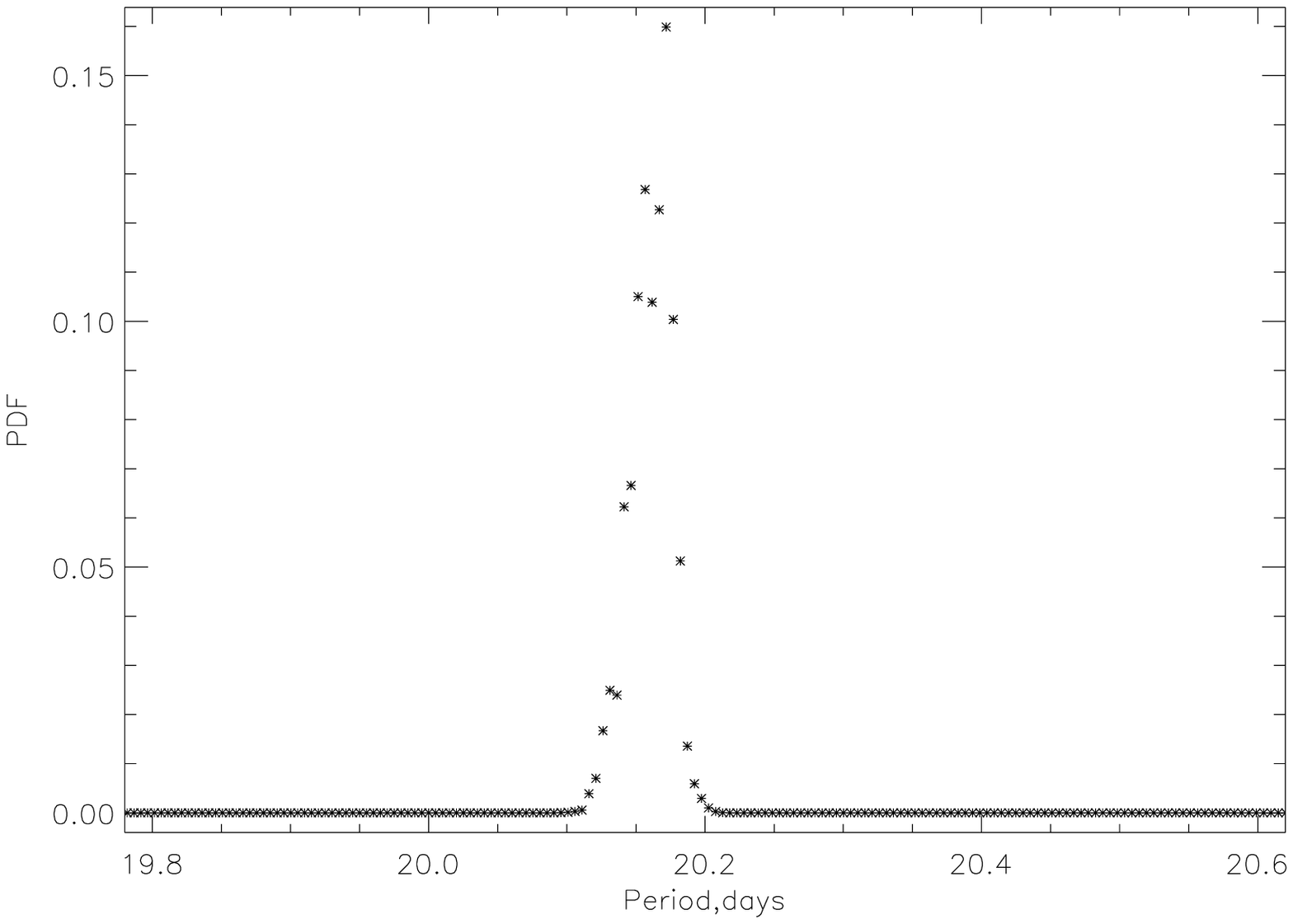,width=7.5cm, height=4.5cm}
\hspace{0.25cm}
\psfig{file=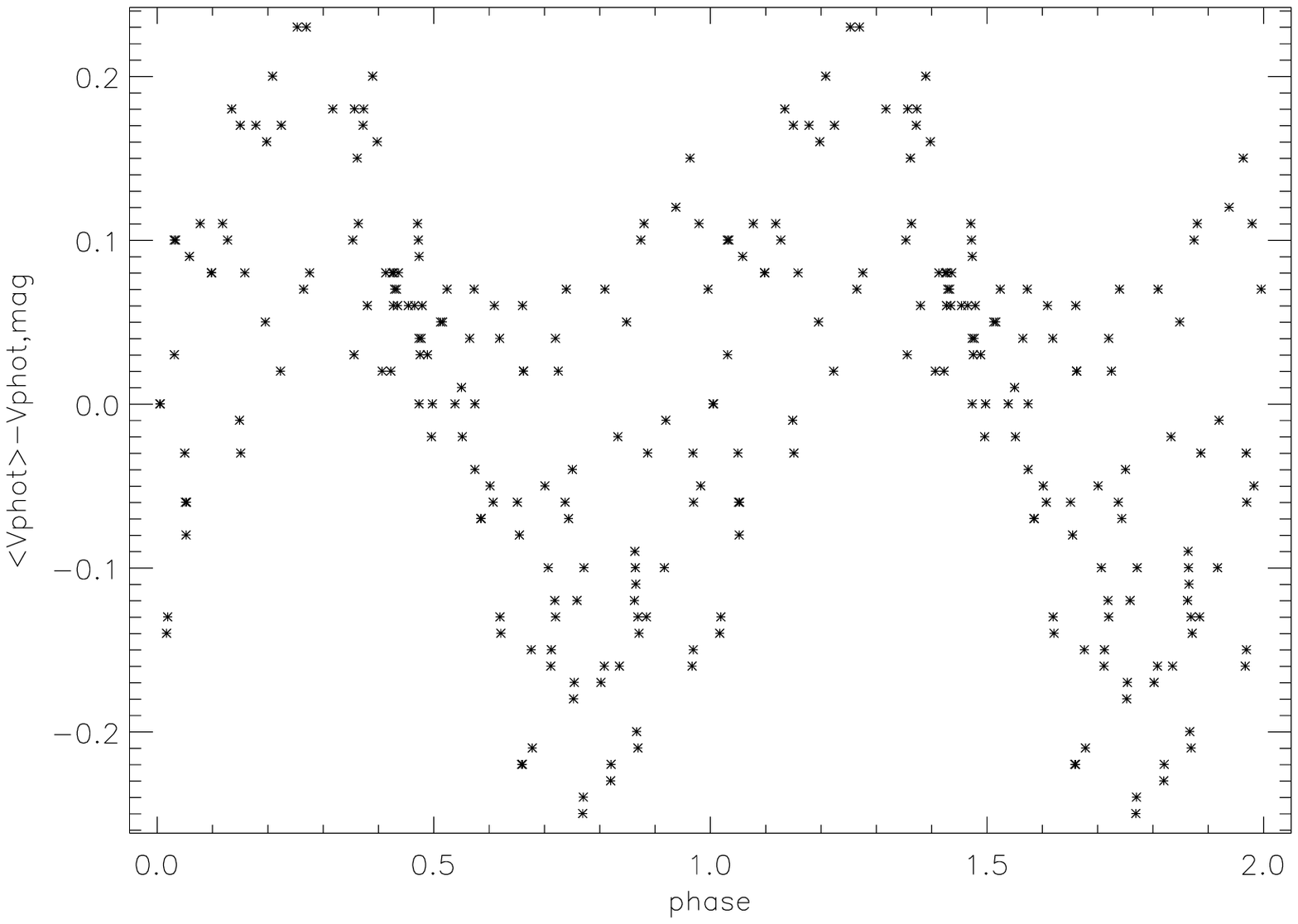,width=7.5cm, height=4.5cm}
}
\hbox{
\psfig{file=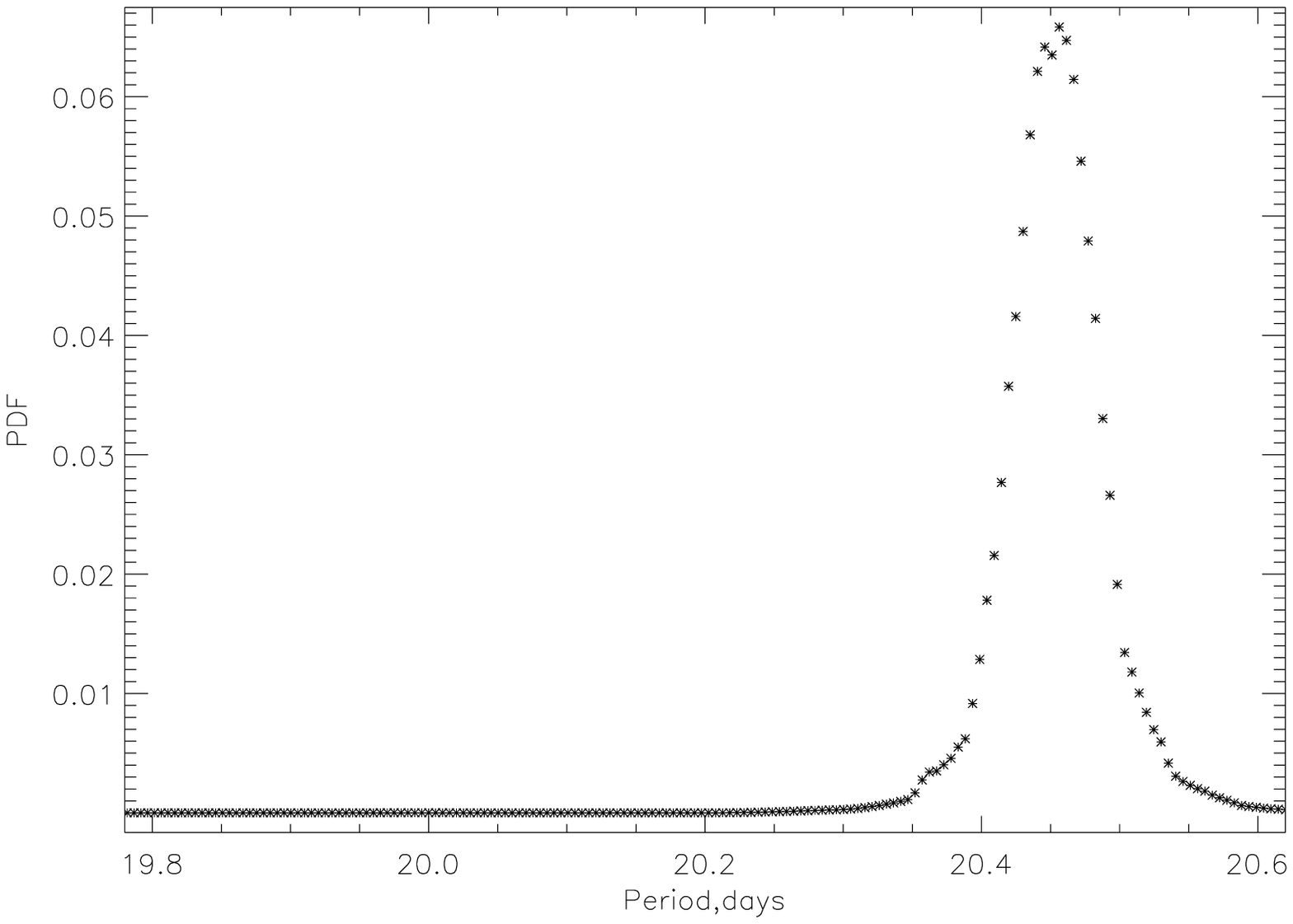,width=7.5cm, height=4.5cm}
\hspace{0.25cm}
\psfig{file=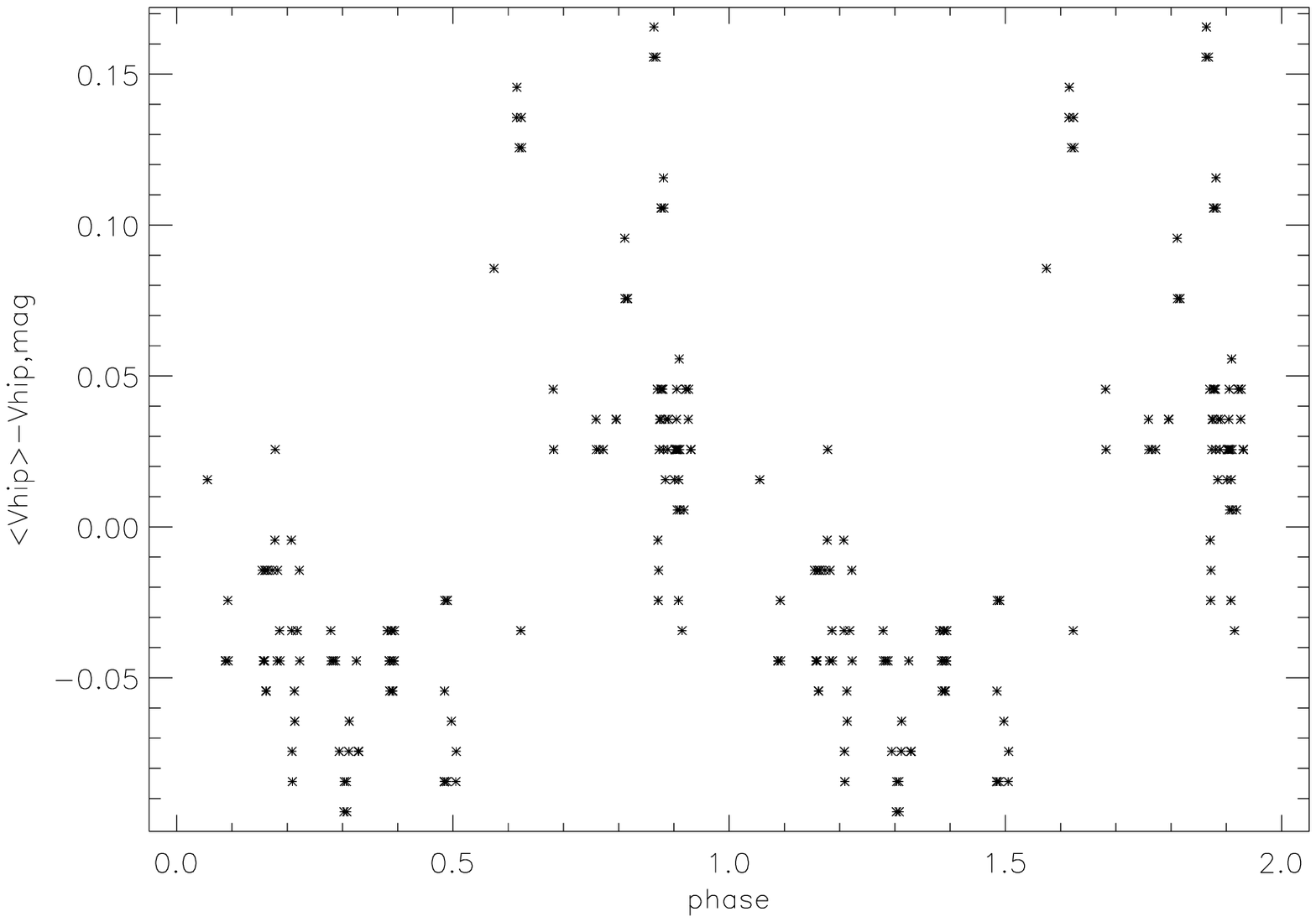,width=7.5cm, height=4.5cm}
}
\hbox{
\psfig{file=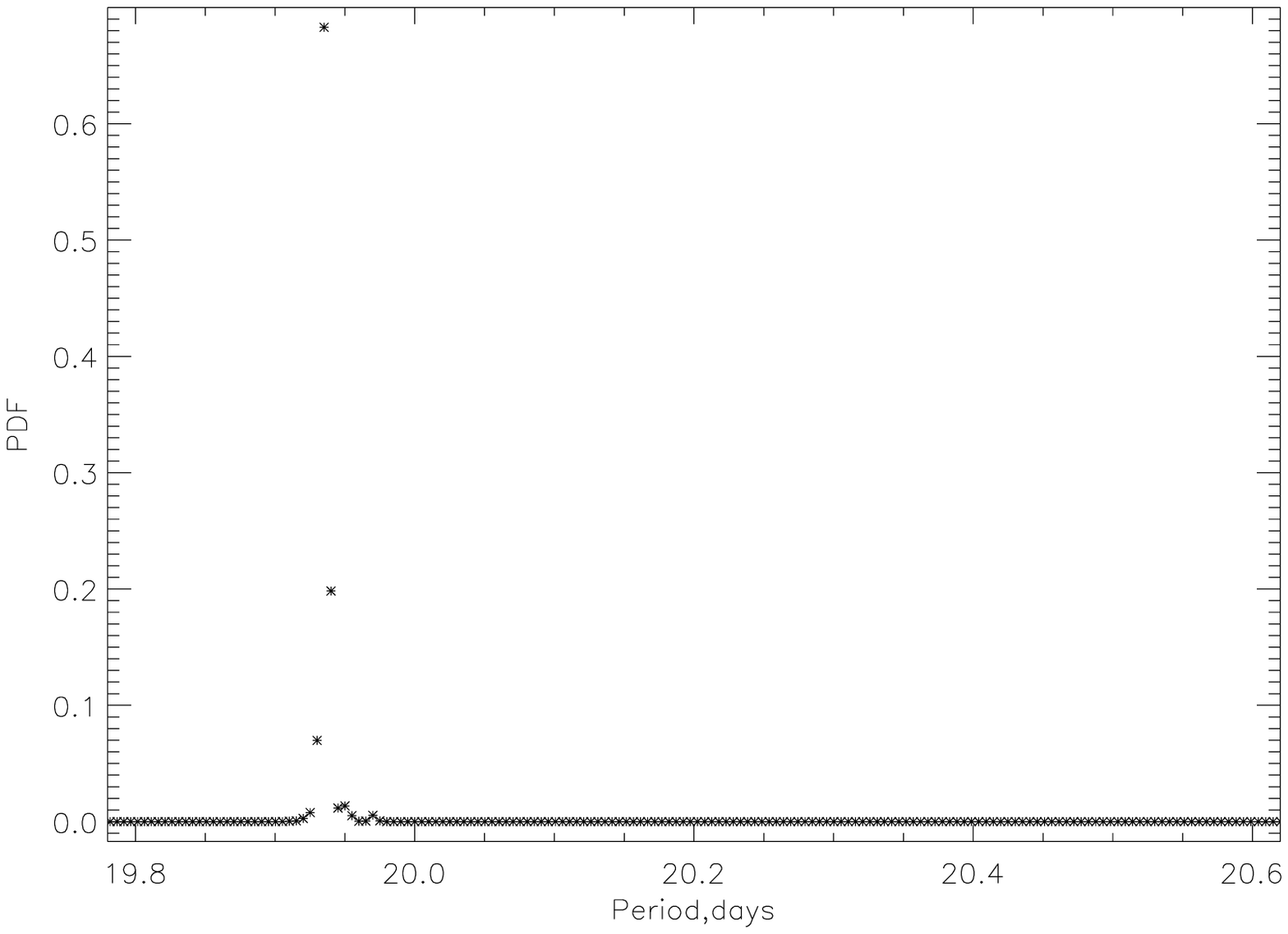,width=7.5cm, height=4.5cm}
\hspace{0.25cm}
\psfig{file=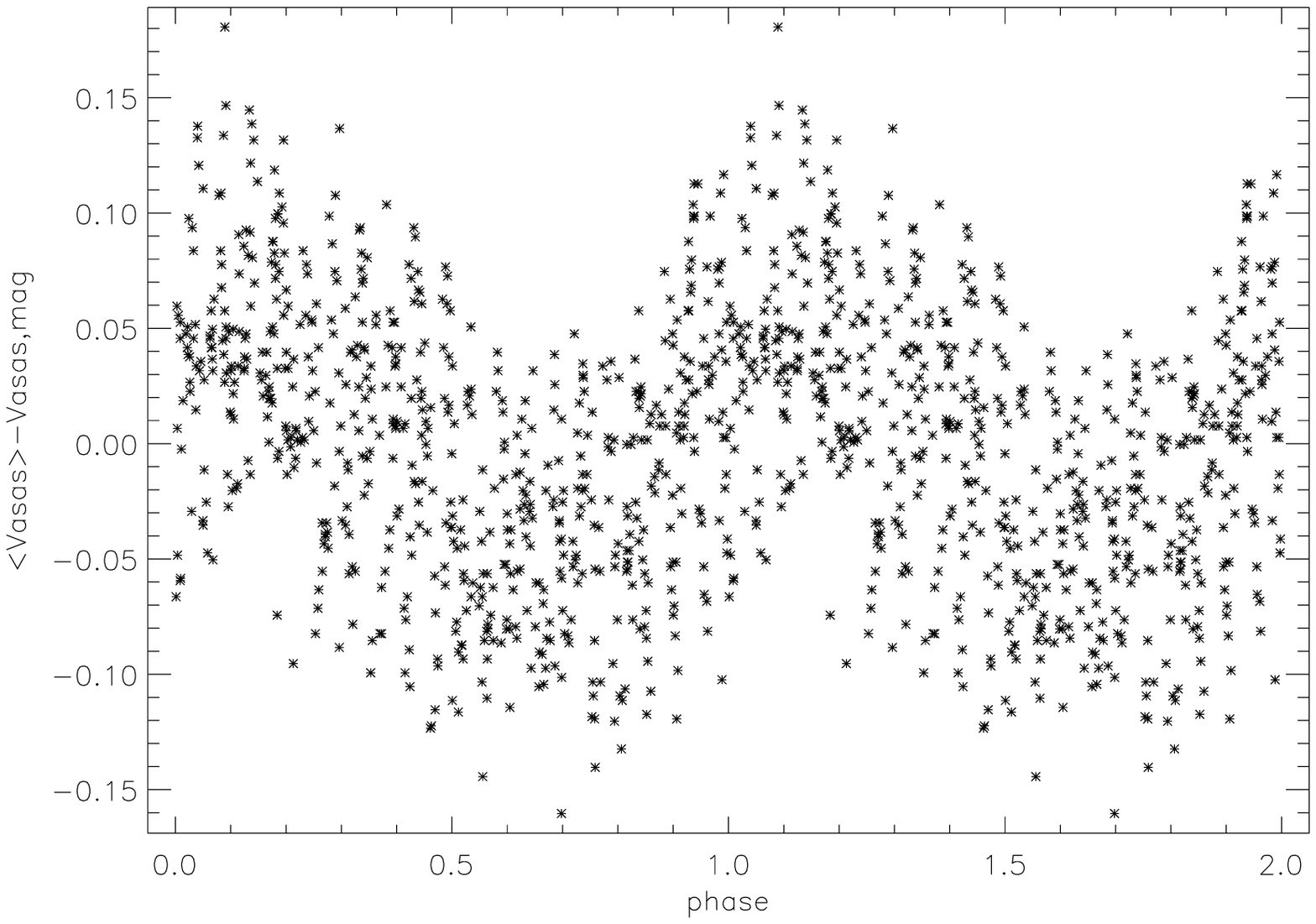,width=7.5cm, height=4.5cm}
}
\hbox{
\psfig{file=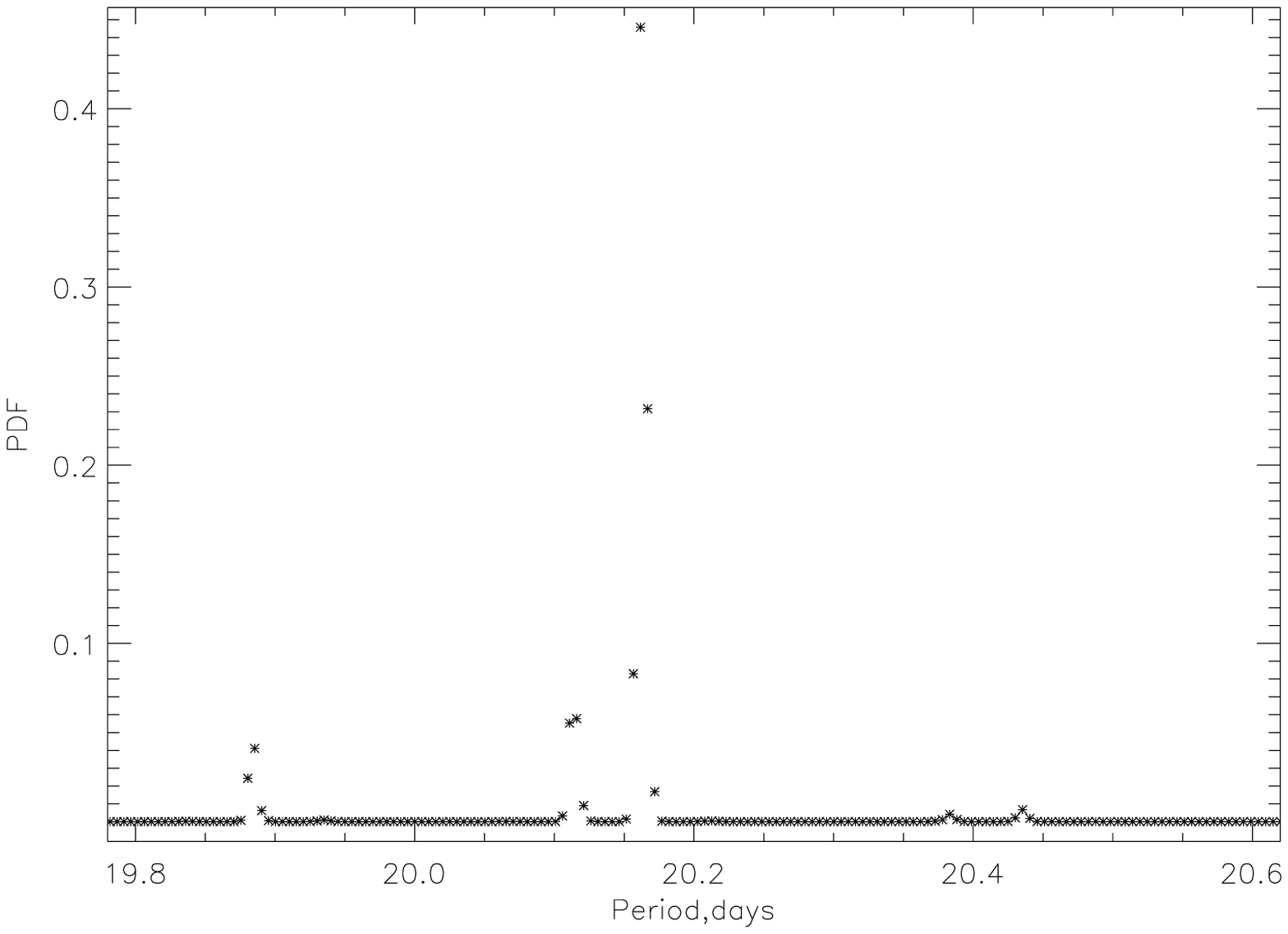,width=7.5cm, height=4.5cm}
\hspace{0.25cm}
\psfig{file=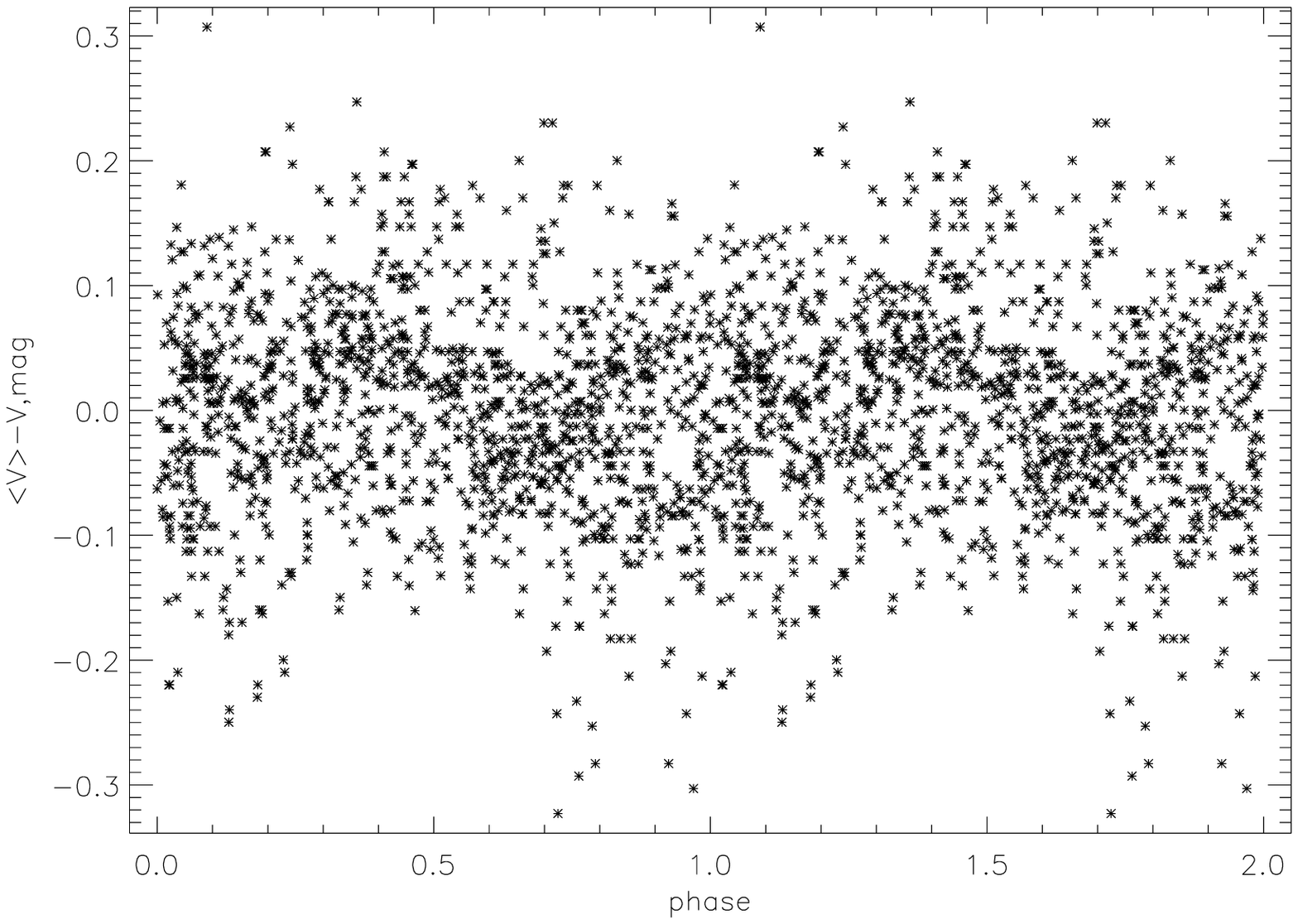,width=7.5cm, height=4.5cm}
}
}
\end{center}

\caption
{Marginal PDFs (see the text) of the period, plots show only the interesting part ({\it left panel}) and mean $V$ subtracted light curve , with common $HJD=2438560.4$ for the zero phase ({\it right panel}) for: from top to bottom - BAM, PHOT, HIP, ASAS and ALL the data.}
\label{Fig. 2}
\end{figure*}
The marginalization of the joint probability function over $m$ represents the information provided by the data for the period. The period that best describes the observed light curve may be estimated by determining the maximum of the marginal posterior probability dencity function, or by calculating the posterior weighted mean period. We have calculated the 68 per cent credible intervals for a given parameter following \citet{GL3} as the interval, that contains 68 per cent of the posterior Probability Density Function (PDF) and where the PDF is everywhere greater than the one outside the credible interval. The Bretthorst generalization of the Lomb-Scargle periodogram gives very close results and also confirms the period variations. Table 2 represents our results: the most probable \textbf{$m_{\rm mode}$}, the most probable noise scale parameter \textbf{$b_{\rm mode}$}, the maximal probable period (\textbf{$P_{\rm mode}$}), the weighted mean period \textbf{$P_{\rm mean}$}, the credible intervals for the period detection and the period derived by Bretthorst generalisation method \textbf{$P_{\rm bret}$}. 
The posterior probabilities are normalized in order to obtain the PDFs and the later have an integral over the priori ranges equal to 1. For the separate datasets as well as for all the data together the expanded plots showing the interesting part of the marginal PDFs are given on the left panels of Fig.~2. The mean subtracted $V$ light curves, plotted with the most probable period for the datasets and for all the data, and with a common HJD for the zero phase for all the photometries, set at the beginning of observations at $HJD=2438560.4$ are presented in the right panels of Fig.~2. For the separate datasets \textbf{$b_{\rm mode}$} as well as the \textbf{$b_{\rm mean}$} are close to $1$, and shows that the periodic model is doing a good representation of the observational data. Our analysis confirms previous suggestions \citep{ib1} for period changes as well as for light curve phase shifting \citep{poll}. The value of the \textbf{$b_{\rm mode}=0.28$} for ALL the data indicates that the simple periodic model does not explain satisfying all the observational data and eventually refers for the presence of additional signal in the data. However, in the case of combined data we do see possible other source of "noise" (on the base of the estimated parameter \textbf{$b$}) and in principle larger credible interval that depends not only from spanned time, but also from the signal to noise ratio and from the amplitude of periodic variations. The mean and mode period obtained for photographic plate and photoelectric photometry data are close to the previous published ones. Scatter in the light curves from different datasets shows that the amplitude of brightness variations, light curve shape and phase of the minimal brightness change with the epoch of observation. This additionally complicates the estimation of the light curve shape which is beyond the goals of the presented research.

\section{Long-term variability of CF Oct}

As it is mentioned above, the GL method gives us opportunity for searching for long-term cycles with duration up to 6000 days. For the proposed study we selected time-scale from 600 to 6000 days that is rather different from the time-scale of the rotational modulation and allows to separately study the long-term variability of CF~Oct. The hypotheses for constant ($H_{\rm C}$), for nonperiodic ($H_{\rm NP}$) and for periodic signal ($H_{\rm P}$) are evaluated, and the odds ($h_{\rm i}/h_{\rm c}$) ratios are given in Table 3. The noise scale parameter for all the models takes into account the rotational modulation as an additional noise. For the nonperiodic model we use number of bins parameter \textbf{$m=40$}. The result shows that the periodic model can best represent the observational data and it is the most probable model. The variable model appears the second probable one and it is very reasonable as periodical models are a special case of variable ones. The noise scale parameters for the examined models give the additional information that the constant and variable models on a long-term scale do not explain significant structures in the data.

\begin{table*}
\centering
\begin{minipage}{140mm}
\caption{The hypotheses space, odds ratios and noise scale parameters}
\begin{tabular}{@{}lcc@{}}
\hline
Hypothesis&odds ratio $h_{\rm i}/h_{\rm 0}$&\textbf{$b_{\rm mode}$}\\
\hline
$H_{C}$ : Constant signal with randon Gaussian noise&$1$&0.21\\
$H_{NP}$ : Nonperiodic signal with unknown shape&$1.8\times10^{2}$&0.54\\
$H_{P}$ : Periodic signal with a priori unknown shape and period&$0.4\times10^{80}$&1.32\\
\hline
\end{tabular}
\end{minipage}
\end{table*}
In order to search for a long-term periodicity we have computed the Bayesian PDF and estimated probabilities for period detection over the mean-subtracted magnitude ($<V>-V$, where $<V>$ is the mean value of $V$ for each data set, i.e. $<\textbf{V}>$=($<V_{\rm bam}>$,$<V_{\rm phot}>$,$<V_{\rm hip}>$,$<V_{\rm asas}>$) data on long-term time scales with a prior period range from 600 to 6000~d. Bayesian estimation of the probability for long-term periodic modulation over mean-subtracted magnitude data results in a marginal PDF (Fig. 3) with three individual peaks with local maxima at $3582$~d ($\sim~9.8$~yr), $2432.5$~d ($\sim 6.7$~yr) and $1173$~d ($\sim 3.2$~yr) and $FWHM$ of $184$~d, $70$~d and $21$~d respectively.


\begin{figure}
\psfig{file=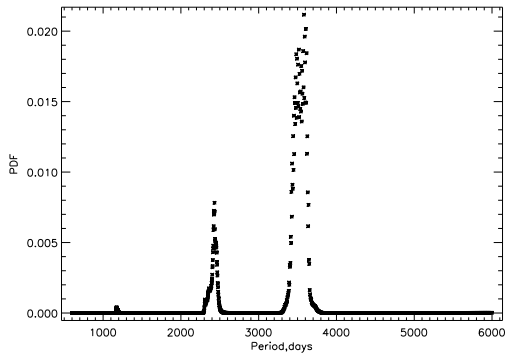,width=7.0cm}
\caption{Marginal PDF of the period for mean-subtracted $V$ magnitude data}
\label{Figure. 3}
\end{figure}

The phase shift of the minimal or maximal light suggested by \citet{poll} and \citet{ib1} has led us to the idea to study time-residuals. The time-residuals represent the difference between the observed and model predicted times of signal minima (or maxima) and are an analog of the $(O-C)$ terms in light curves of the binary stars. They are representing the observed phase shift in the light curve of CF~Oct and are probably connected with the spot migration over the stellar equator. We have computed the time-residuals by using stepwise (with \textbf{$m=10$} bins) light curves of the PHOT dataset as a reference for expected times of minimal brightness of the star. The differences between the expected or predicted by the PHOT light curve and actually observed times of minimum brightness for all the observed data span were calculated. The marginalized (over \textbf{$m$}) probabilities for detection of cyclic activity with period in the interval $600$ to $6000$~d are presented on Fig.~4. The marginal total probability of 0.46 for cycle with a period of $2603.0$~d ($\sim 7.1$~yr) with $FWHM$ of $54$~d is estimated. This cycle length is in a good agreement with the suggestions of \citet{poll}. A period of $2648$~d was detected with the Bretthorst generalization method and falls very close to the period determined by GL method. Fitting the time-residual data with a stepwise curve with period of $2603.0$~d is presented on Fig. 6. 

The other two peaks of the marginal PDF with periods near $4100$ and $5500$~d and $FWHM$ of $432$ and $769$~d have relatively low probabilities both of $0.27$ and they give much more scatter of the stepwise fitting curve compared to the $2603.0$~d period. We suppose that the minor peaks are probably due to the data gaps and the quasi-random data distribution. Thus we consider that the period of $2603.0$~d is the most probable for the time-residual data.

\begin{figure}
\psfig{file=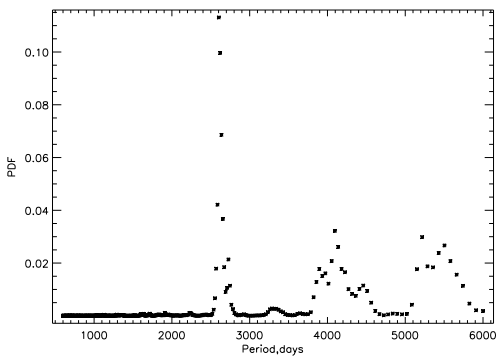,width=7.0cm}
\caption{Marginal PDF of the period for time-residual data}
\label{Figure. 4}
\end{figure}

\section{Discussion}
The results from Bayesian detection of the rotational modulation for CF~Oct are in good agreement with previously published values. Using the Bayesian statistical method meant we were able to estimate the probabilities for rotational modulation over the priori selected range, based on the previous studies, and to determine the mode and mean periods derived from separate datasets as well as for all the observational data. The rather narrow credible intervals for period detection over the four separate datasets, state the range where the simple periodic model best describes the observations and thus we propose that the star probably exhibit variations in the rotational period. We have tested the hypotheses for constant, simple periodic and variable model for all the observed data and find out that the periodic model is still the most probable one. Obviously this model does not explain the changes in the rotational period, so more complicated analysis with expanded hypotheses space to account for both rotational and long-term activity of CF~Oct need to be performed. We were able also to estimate the mode period of $20.16$~d for all the collected data, but with relatively large credible interval. This is very close to the best period estimated over electrophotometric data by \citet{poll}. The estimated rotational period varies from $19.94$~d for ASAS data to $20.46$~d for HIP data. The most evident hypothesis for changes in the rotational period is differential rotation of the stellar surface, and latitude migration of surface spots. It is noteworthy to point out that rapid changes in the light curve of CF~Oct were observed by \citet{innis3} in 2006. ASAS data set does not have enough observations at the time of these changes that would give additional information. However for the ASAS observations after the reported changes near $HJD=2454040.0$ the amplitude of the variation is relatively small, about $0.2$~mag.

\citet{poll} suggested for activity cycle of $9\pm3$ years based on the changes of the amplitude of the spot-wave. The period for cyclic activity of $2603$~d estimated with GL method by the time-residual data is in an agreement with this prediction. This new found cycle is another indication for changes in the rotational period of the star. 
Inspection of the stepwise light curve with period $2603$~d and (Fig. 6) overploted on the time-residual data shows, that characteristic features - minima and maxima of the light curve are well supported by the observations. 
The derived period for cyclic activity of $2432.5$~d over the mean-subtracted magnitude data is close to the period derived from time-residual data, but the $FWHMs$ of these cycles does not overlap. Thus the hypothesis of the existence of the joint period for the magnitude and time-residual cycles in the presence of rotational modulation is very interesting and will be tested later by the use of the method used by \cite{GL2}. The estimated cycle of $3582~d$ (Fig. 5) by the use of magnitude data according the represented parameters is the most probable one, and also fits into the suggested cyclic activity by \citet{poll}.

The $\log_{10}(P_{\rm rot}/P_{\rm cyc})=-2.1$ index for CF Oct, for $P_{\rm cyc}=2603$~d, is in a good agreement with the values for active stars summarised by \citet{olah1}, nevertheless on fig. 13 from the mentioned paper CF~Oct would lie very close to the similar RS~CVn star HK~Lac and would be also close to the line of the main sequence stars of \citet{bal}. On the basis of other detections of long-term variations with two to three cycles for several stars reported by \citet{olah1}, and for the activity cycles analysis of \citet{fro} for HK~Lac, we also propose that the hypotheses for complex long-term variability with superposition of two or more cycles of CF~Oct are relevant and needs future evaluating by the means of Bayesian statistics.

\begin{figure}
\psfig{file=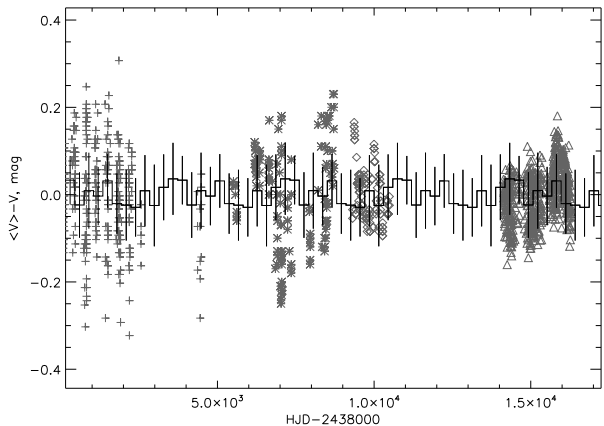,width=8.5cm}
\caption{Stepwise, phase-shifted light curve for the cycle with period of $3582$~d plotted over the mean-subtracted $V$ magnitude data in gray colour. Symbols are the same as on Fig. 1.}

\label{Figure. 5}
\end{figure}

\begin{figure}
\psfig{file=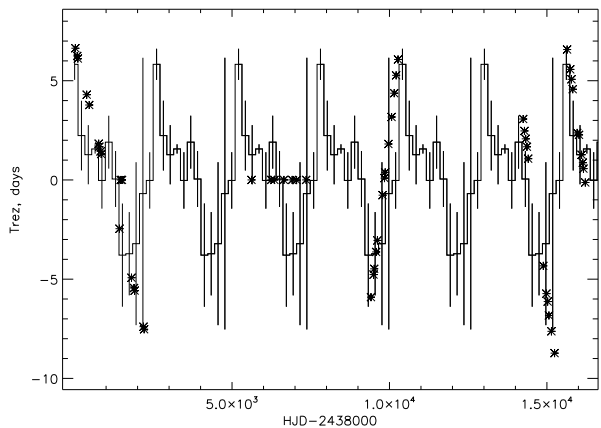,width=8.5cm}
\caption{Stepwise time-residual fitting curve for the cycle with period of $2603~d$ overploted over the time-residual ($Trez$) data. }

\label{Figure. 6}
\end{figure}

\section{Conclusions}
The main result of our work is that it clearly shows the evidence of a long-term activity cycle of the active giant star CF~Oct. Two cycles of $2603$~d and $3582$~d with $FWHMs$ of $53$ and $184$~d were detected by the use of two different statistical techniques. First one cycle of $2603$~d represent the changes of stellar magnitude, while the second one is related to the changes of the phase of minimal brightness of the star, so it means that there is no superposition of the two cycles.

We have given arguments that support previous suggestions for smooth period changes, due to differential rotation and possible spot migration, and derived the most probable periods for different intervals of the light history of CF~Oct. The estimated rotational period are: $20.04$~d for the interval 1964--76 from Bamberg observatory plate data, $20.16$~d for the interval 1979--88 from electrophotometry data, $20.45$~d for the interval 1989--93 from Hipparcos data and $19.94$~d for the interval 2000--09 from ASAS archive.
The Bayesian analysis indicates that the hypotheses for presence complex cyclic activity of CF~Oct is also relevant and it is a subject of future work to evaluate the proposition for observation of several long-period cycles over the joint analysis of magnitude and time-residual data.

This work demonstrates the value of the astronomical data archives and draws attention to preserving archival observations for future exploitation. With the extension of the observational time series by the photographic plate data we are able to study decadal variations of bright active stars.

\section*{Acknowledgments}

We are thankful to Prof. Ralph Neuh\"auser and the support of AIU-Jena for hosting the present statistical study; Dr.
B. Pilechki for the help and for providing us additional information regarding ASAS data;
Prof. U. Heber and AvH foundation for supporting our work for digitizing plates in use from BOSSS. This work is also performed with the support of Bulgarian NSF, grant DO 02-275.

\bsp

\label{lastpage}

\end{document}